\documentclass[reprint,superscriptaddress,bibnotes, amsmath,amssymb,prb]{revtex4-1}

\usepackage{graphicx}
\usepackage{dcolumn}
\usepackage{bm}

\begin{document}

\title{Correlation between non-Fermi-liquid behavior and superconductivity in
    (Ca, La)(Fe, Co)As$_2$ iron arsenides: A high-pressure study}

\author{W. Zhou}
\affiliation{Advanced Functional Materials Lab and Department of Physics, Changshu Institute of Technology, Changshu 215500, China}
\affiliation{School of Physics and Key Laboratory of MEMS of the Ministry of Education, Southeast University, Nanjing 211189, China}
\author{F. Ke}
\affiliation{Center for High Pressure Science and Technology Advanced Research, Shanghai 201203, China}
\author{Xiaofeng Xu}
\email{xiaofeng.xu@cslg.edu.cn}
\affiliation{Advanced Functional Materials Lab and Department of Physics, Changshu Institute of Technology, Changshu 215500, China}
\author{R. Sankar}
\affiliation{Institute of Physics, Academia Sinica, Nankang, Taipei R.O.C. Taiwan 11529}
\affiliation{Center for Condensed Matter Sciences, National Taiwan University, Taipei 10617, Taiwan}
\author{X. Xing}
\affiliation{School of Physics and Key Laboratory of MEMS of the Ministry of Education, Southeast University, Nanjing 211189, China}
\author{C. Q. Xu}
\affiliation{Advanced Functional Materials Lab and Department of Physics, Changshu Institute of Technology, Changshu 215500, China}
\author{X. F. Jiang}
\affiliation{Advanced Functional Materials Lab and Department of Physics, Changshu Institute of Technology, Changshu 215500, China}
\author{B. Qian}
\email{njqb@cslg.edu.cn}
\affiliation{Advanced Functional Materials Lab and Department of Physics, Changshu Institute of Technology, Changshu 215500, China}
\author{N. Zhou}
\affiliation{School of Physics and Key Laboratory of MEMS of the Ministry of Education, Southeast University, Nanjing 211189, China}
\author{Y. Zhang}
\affiliation{School of Physics and Key Laboratory of MEMS of the Ministry of Education, Southeast University, Nanjing 211189, China}
\author{M. Xu}
\affiliation{School of Physics and Key Laboratory of MEMS of the Ministry of Education, Southeast University, Nanjing 211189, China}
\author{B. Li}
\affiliation{College of Science, Nanjing University of Posts and Telecommunications, Nanjing 210023, China}
\author{B. Chen}
\affiliation{Center for High Pressure Science and Technology Advanced Research, Shanghai 201203, China}
\author{Z. X. Shi}
\email{zxshi@seu.edu.cn}
\affiliation{School of Physics and Key Laboratory of MEMS of the Ministry of Education, Southeast University, Nanjing 211189, China}

\date{\today}

\begin{abstract}
Non-Fermi-liquid (NFL) phenomena associated with correlation effects have been widely observed in the phase diagrams of unconventional superconducting families. Exploration of the correlation between the normal state NFL, regardless of its microscopic origins, and the superconductivity has been argued as a key to unveiling the mystery of high-$T_c$ pairing mechanism. Here we systematically investigate the pressure-dependent in-plane resistivity ($\rho$) and Hall coefficient ($R_H$) of a high quality 112-type Fe-based superconductor Ca$_{1-x}$La$_x$Fe$_{1-y}$Co$_y$As$_2$ ($x= 0.2$, $y= 0.02$). With increasing pressure, the normal state resistivity of the studied sample exhibits a pronounced crossover from non-Fermi-liquid to Fermi-liquid (FL) behaviors. Accompanied with this crossover, $T_c$ is gradually suppressed. In parallel, the extremum in Hall coefficient $R_H(T)$ curve, possibly due to anisotropic scattering induced by spin fluctuations, is also gradually suppressed. The symbiosis of NFL and superconductivity implies that these two phenomena are intimately related. Further study on the pressure-dependent upper critical field reveals that the two-band effects are also gradually weakened with increasing pressure and reduced to one-band Werthamer-Helfand-Hohenberg limit in the low $T_c$ regime. Overall, our study supports the picture that NFL, multigap, extreme $R_H(T)$ are all of the same magnetic origin, i.e., the spin fluctuations in the 112 iron arsenide superconductors.
\end{abstract}


\maketitle

\section{Introduction}
Exploring the unusual normal state of high temperature (high-$T_c$) superconductors is arguably one of the most important tools to pry into the high-$T_c$ superconductivity mechanism \cite{Ando,Dagan,Daou,Analytis,Dai1}. In the two largest high-$T_c$ families, namely cuprates and iron-based superconductors, a significant deviation from the quadratic $T$-dependence of resistivity that is expected from Landau's Fermi-liquid (FL) theory has been found just above the superconducting dome, revealing the close tie between non-Fermi-liquid (NFL) excitations and the high-$T_c$ superconductivity \cite{Ando,Analytis}. Besides the NFL phenomenon, iron-based superconductors also display other extraordinary transport behaviors, such as a strong $T$-dependent Hall coefficient $R_H$ \cite{Fanfarillo,Ding}, sign changes in $R_H(T)$ curves \cite{Ohgushi,Zhuang} and the violation of Kohler's scaling \cite{Cheng,Kasahara}. As a consequence of the close proximity between the antiferromagnetic (AFM) order and superconductivity in the phase diagram, a very plausible origin for these anomalous transport properties is the spin fluctuation (SF) which may also serve as the pairing glue for the high-$T_c$ superconductivity.

In iron pnictides, the normal-state resistivity has often been extensively described by the power-law dependence, $\rho=\rho_0+AT^\alpha$, in which $\alpha = 2$ corresponds to Fermi-liquid (FL) ground states \cite{Jiangshuai,Dai,Kasahara1,Eom}. Deviations from FL ($1<\alpha<2$) have been observed in both the under- and over- doped regions, revealing a remarkable V-shaped dependence of the exponent $\alpha$ on doping \cite{Kasahara1,Eom}. The observation of $\alpha = 1$ at the optimal doping seemingly support the magnetic quantum critical point (QCP) scenario. However, there is also a lot of evidence showing that NFL behaviors could be a new state of matter in its own right rather than a consequence of the QCP. For example, in Ce$_{1-x}$Yb$_x$CoIn$_5$, NFL regimes and superconductivity persist in samples with and without QCP \cite{Hu}. In Co-doped LiFeAs, which is regarded as a simple system because neither magnetic nor structural transitions have been detected in the phase diagram, the NFL region is found to shift to the boundary of the superconducting phase \cite{Dai}. Currently, the relationship between the normal-state NFL, QCP and superconductivity, especially in connection with SFs, is still a matter of considerable controversy.

\begin{figure}
\includegraphics[width=9.8cm]{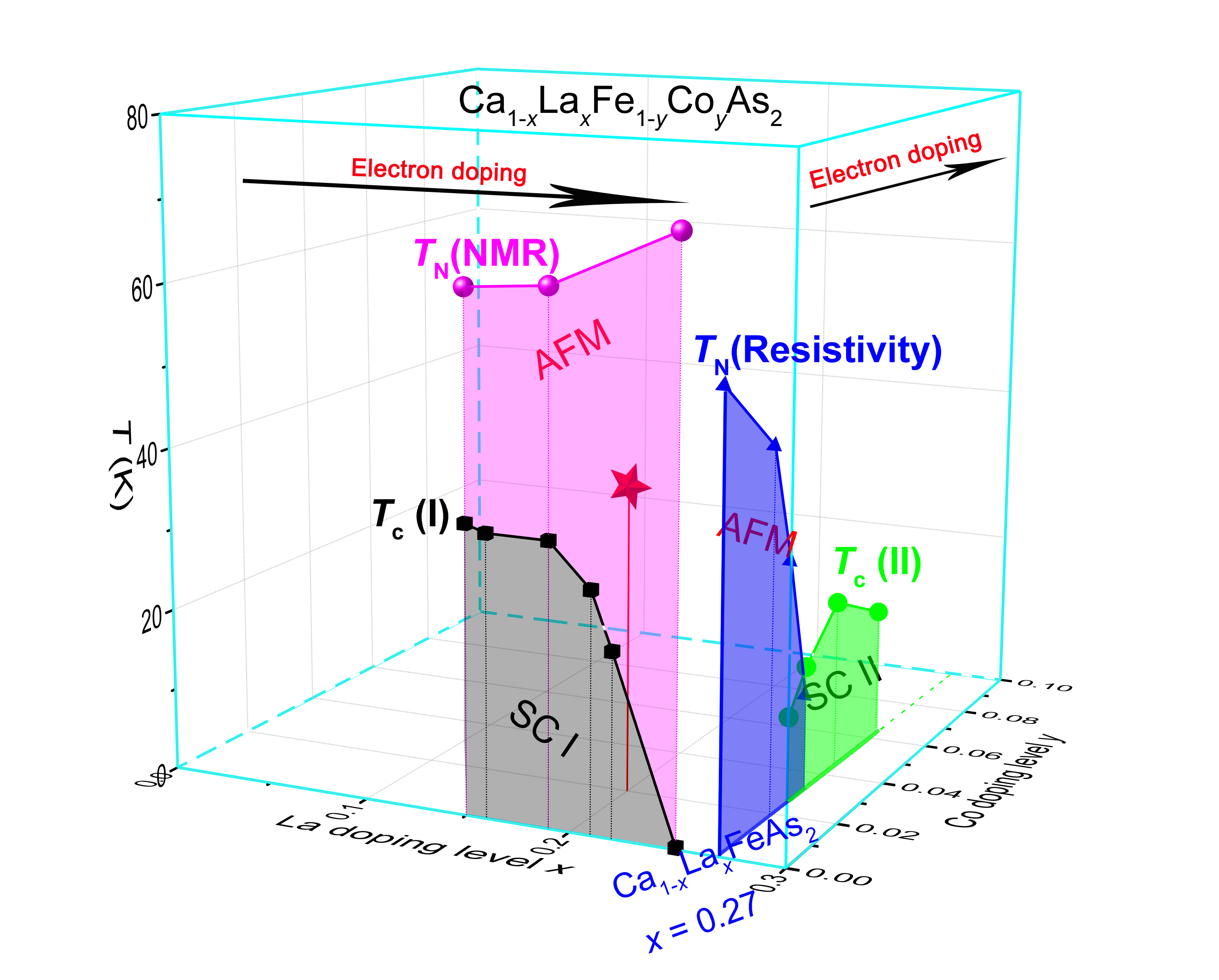}
\caption{\label{phasediagram} The electronic phase diagram for 112-type Fe-based superconductor Ca$_{1-x}$La$_x$Fe$_{1-y}$Co$_y$As$_2$ based on previous experimental results \cite{Kawasaki,Jiang1}. For Ca$_{1-x}$La$_x$FeAs$_2$, the AFM transition temperature $T_N$ grows with increasing electron doping from La substitution on the Ca site. While, for Ca$_{1-x}$La$_x$Fe$_{1-y}$Co$_y$As$_2$ ($x$ = 0.27), $T_N$ is gradually suppressed with electron doping from Co substitution on the Fe site. The sample labelled by the red star is the 112 superconductor Ca$_{1-x}$La$_x$Fe$_{1-y}$Co$_y$As$_2$ ($x = 0.2$, $y = 0.02$, $T_c \sim 37.5$ K at ambient pressure) used for the high pressure study in this work.}
\end{figure}

Recently, novel 112-type superconductors Ca$_{1-x}$RE$_x$FeAs$_2$ (RE = rare earth element) were discovered and provided a new
platform to study the relationship between NFL and unconventional superconductivity \cite{Katayama,Yakita}. For the 112 system, it is theoretically predicted that the parent phase CaFeAs$_2$ is a spin-density-wave (SDW)-type striped antiferromagnet driven by Fermi surface nesting \cite{Huang}, similar to many other Fe-based superconductors. However, no experimental synthesis of pure CaFeAs$_2$ has yet been achieved. As shown in the summarized phase diagram in Fig. 1, doping CaFeAs$_2$ with La can stabilize a high-$T_c$ superconductivity over 35 K \cite{Katayama,Kawasaki}. For samples with La doping level $0.15 \leq x \leq 0.24$, AFM-like transitions were claimed from $^{75}$As-nuclear magnetic resonance (NMR), and the N\'{e}el temperature $T_N$ was intriguingly found to increase with $x$ \cite{Kawasaki}, in stark contrast to many existing Fe-based superconductors. Later, a lower $T_N$ ($\sim$58 K) was observed from the resistivity anomaly at $x = 0.27$ by another group \cite{Jiang}. More intriguingly, if one further introduces electron doping through Cobalt-co-doping for Fe or application of high pressure on sample Ca$_{1-x}$La$_x$FeAs$_2$ ($x = 0.27$), AFM transition is found to be gradually suppressed and a seemingly second superconducting phase with a maximum $T_c$ around 25 K is resolved \cite{Jiang1,Zhouyazhou}. It appears that, sample Ca$_{1-x}$La$_x$FeAs$_2$ ($x = 0.27$) plays the role of the '\textit{parent phase}' for the second superconducting regime. Such an interesting phase diagram with multiple electronic orders is rarely seen in other high-$T_c$ superconducting systems, making the 112-type superconductors a unique platform to study the interplay between the normal-state transport and superconductivity.

In this work, we have selected the high-quality single crystal Ca$_{1-x}$La$_x$Fe$_{1-y}$Co$_y$As$_2$ ($x= 0.2$, $y= 0.02$, $T_c \sim 37.5$ K), whose normal state exhibits pronounced departure from the Fermi-liquid picture \cite{Xing1}, as a candidate to perform the high pressure study. Compared with chemical substitutions, high pressure has proven to be a clean way to tune the electronic properties of a material as no additional impurity effects are introduced. Unfortunately, high-pressure study of the correlation between the normal state transport and high-$T_c$ superconductivity on 112 iron pnictides is still lacking. In this work, we aim to achieve a fine tuning of the normal-state transport and superconductivity via pressure, and thus investigate the correlation between NFL and superconductivity in this 112 iron pnictide. As we found, at low pressures, the normal state of the sample exhibits obvious deviation from the conventional Fermi-liquid (FL) behaviors. Especially, $T$-linear resistivity without signature of possible QCP is observed at the optimal pressure ($P$= 3.9 GPa) where $T_c$ reaches the highest value. Above $P = 17$ GPa where superconductivity is totally suppressed, the normal-state FL behaviors emerge. No trace of a second superconducting phase can be identified up to 40 GPa. We discussed our results in terms of the role of SFs on both the normal-state transport and superconductivity in this 112 Fe-based system.

\section{Experiment}

\begin{figure*}
\includegraphics[width=15cm]{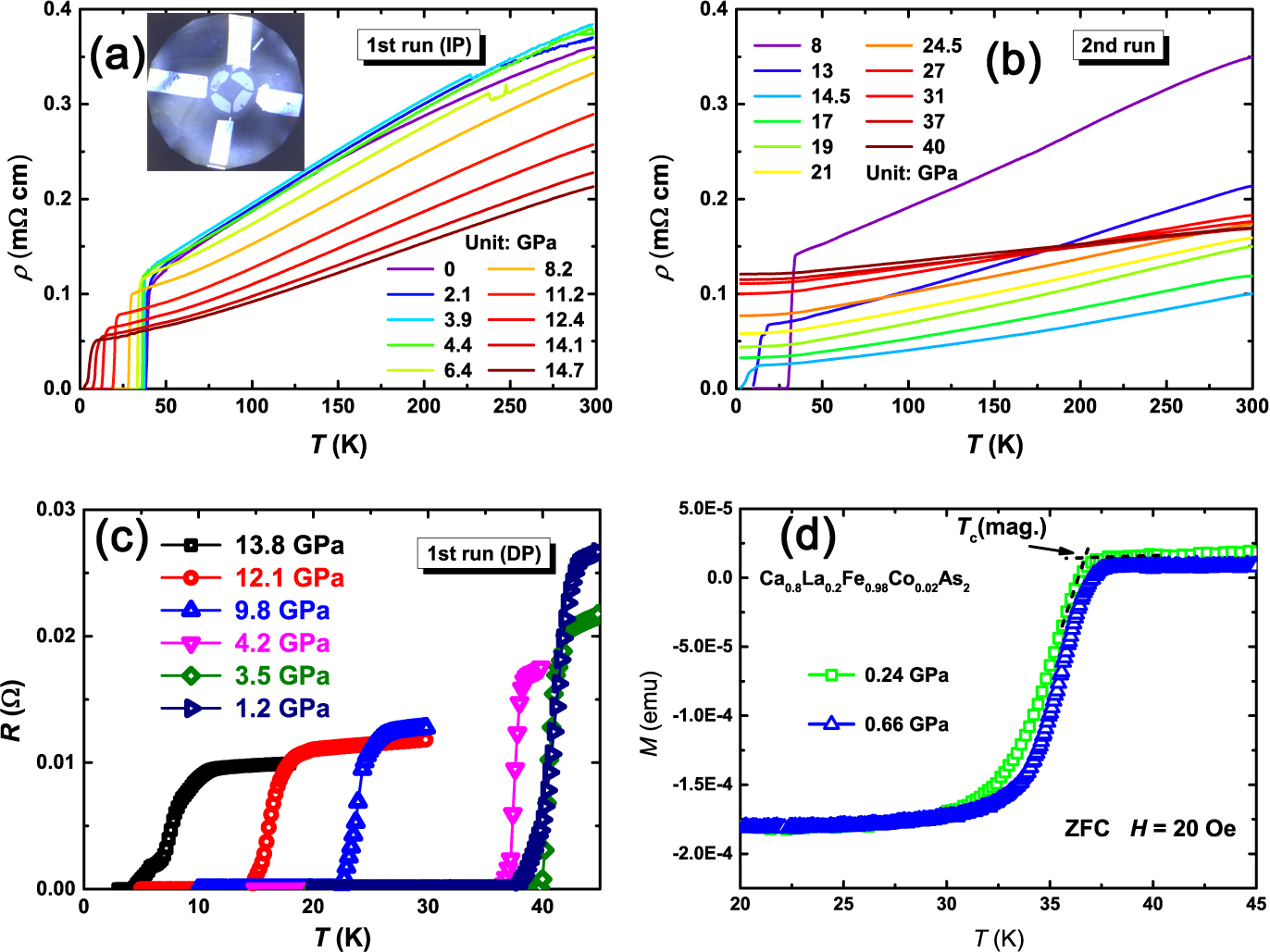}
\caption{\label{RT} (a) and (b) Temperature dependence of resistivity under different pressures during the pressure increasing (IP) process. Inset shows the photograph of the sample and the electrical contacts inside the pressure chamber. (c) Temperature dependence of resistance under different pressures in the pressure releasing (DP) process. Superconductivity is found to be reversibly recovered. (d) Temperature dependence of magnetization of Ca$_{0.8}$La$_{0.2}$Fe$_{0.98}$Co$_{0.02}$As$_2$ under different pressures. The measurements were performed via a zero-field-cooling (ZFC) process under a magnetic field of $H =$ 20 Oe.}
\end{figure*}

Single crystals of Ca$_{1-x}$La$_x$Fe$_{1-y}$Co$_y$As$_2$ ($x= 0.2$, $y= 0.02$) were grown by FeAs self-flux method as described elsewhere\cite{Zhou,Xing}. Good single crystallinity of the as-grown samples was confirmed by a single-crystal x-ray diffraction measurement \cite{Xing}. The electrical pressure experiments were performed using a screw-type diamond anvil cell. A pair of anvil culets of 500 $\mu$m was used at the top and the bottom of the pressure chamber whose diameter is about 100 $\mu$m. Four electrodes were made on the surface of the freshly-cleaved crystal of size about 70 $\mu$m in length or width, and 15 $\mu$m in thickness (see the pressure set-up in the upper-left inset in Fig. \ref{RT} (a)). Afterward, the sample together with ruby spheres for pressure calibration was loaded into the center of the pressure chamber. The transmitting medium (Daphne 7373) was injected to ensure a quasi-isotropic pressure environment. For the discussion of pressure uniformity issue, see Ref. \cite{SI}. Pressure was calibrated by the measurements of the peak shift in the Raman spectroscopy (laser wavelength $\sim$ 514.5 nm) of the ruby spheres or the diamond anvil. Usually, the pressure was calibrated twice, i.e., before and after each transport measurement. Then the average pressure was taken in the figures.

The dc electrical transport measurements were carried out in a physical property measurement system (PPMS-9 T, Quantum Design). For $\rho(T)$ measurements, the electrodes from the same side of the rectangular sample were set as current source or voltage output. To get precise measurement results, we used a constant temperature sweeping rate of 0.5 K/min across the superconducting transitions. For temperatures above $T_c$, the temperature sweeping rate is set as 1 K/min for all the measurements under different pressures, which enables the accurate analysis of the normal-state transport behavior without the contamination from inconsistent measurement conditions. For Hall measurements, two pairs of diagonal electrodes were individually set as the current source and the Hall voltage output. Two different measurement strategies, namely, field sweeping and temperature sweeping methods, were adopted, and the measurement results are consistent. To cancel the electrode asymmetric factor, Hall resistivity is calculated via $\rho_{xy} = [\rho_{xy}(\mu_0H > 0) - \rho_{xy(}\mu_0H < 0)]/2$. Low-field high-pressure magnetization experiment was performed using a vibrating sample magnetometer (VSM) integrated in PPMS on the basis of a commercial, HMD, Be-Cu piston-cylinder pressure cell. Daphne 7373 was also used as the pressure transmitting medium. Superconducting Pb sample was used as a pressure gauge by measuring its pressure-dependent $T_c$.

\section{Results}

\begin{figure*}
\includegraphics[width=14cm]{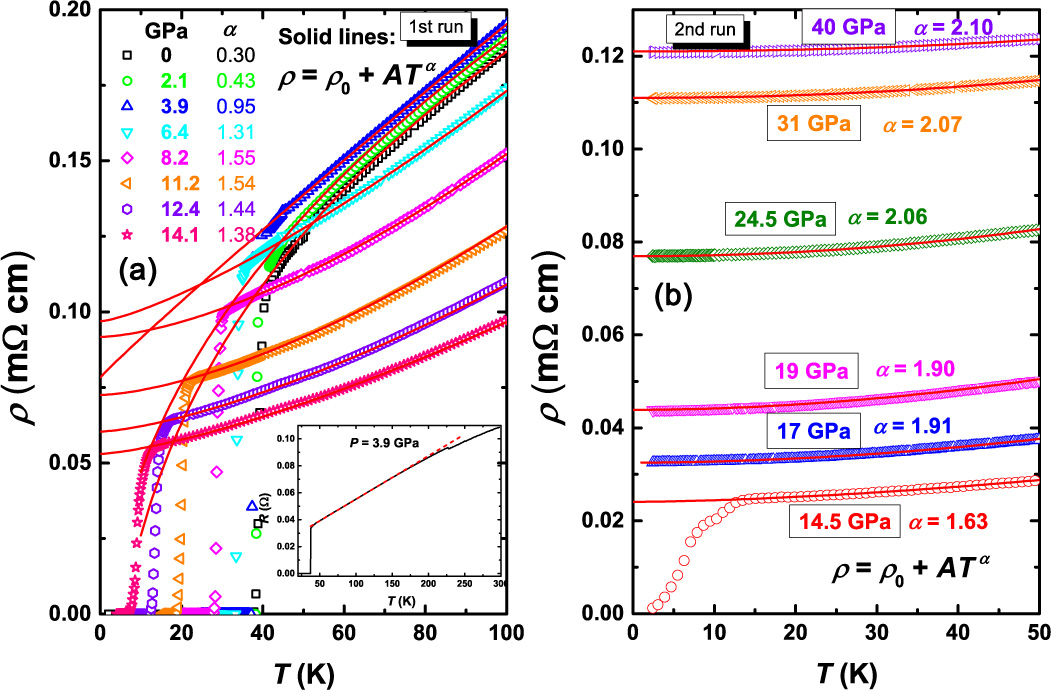}
\caption{\label{Powerlaw} (a) and (b) Power-law ($\rho(T) = \rho_0 + AT^\alpha$) fitting of the normal-state resistivity above $T_c$. Inset in (a) shows the linear $\rho(T)$ behavior within a wide temperature range at $P = 3.9$ GPa. The red dashed line is the guide to eyes and the jump at $\sim$ 230 K is the measurement noise.}
\end{figure*}

As a small amount of Co-doping can greatly improve the superconducting properties of the 112 system and results in a homogenous bulk superconductivity \cite{Xing1,Xing}, sample Ca$_{1-x}$La$_x$Fe$_{1-y}$Co$_y$As$_2$ ($x= 0.2$, $y= 0.02$) was selected as the candidate for the high pressure experiments. Figure \ref{RT} presents the main high pressure experiment results. The sample shows a sharp superconducting transition around 37.5 K at ambient pressure. In the normal state, a downward curvature is evident in the $\rho(T)$ curve, which resembles those of Co-free 112-type crystals \cite{Zhou}, NdFeAsO$_{1-x}$F$_x$ \cite{Cheng}, and over-doped Ba$_{1-x}$K$_x$Fe$_2$As$_2$ \cite{Shen}. Upon applying pressure, the superconducting transition $T_c$ shows somewhat subtle enhancement, which is further verified by the pressure-dependent magnetization measurements as shown in Fig. \ref{RT} (d). This slight positive pressure dependence is analogous to what was seen in the optimally doped Ba$_{1-x}$K$_x$Fe$_2$As$_2$ films \cite{Park}. At low pressure range of $P <$ 3.9 GPa, the normal-state resistivity changes a little with varying pressures, and the downward curvature of normal-state $\rho(T)$ is gradually suppressed. At $P \sim 3.9$ GPa, $T_c$ reaches the highest value while the resistivity above $T_c$ grows linearly with temperature. As $P$ increases over 3.9 GPa, $T_c$ begins to decrease with pressure. Simultaneously, an upward curvature begins to be evident in an intermediate temperature range from $T_c$ to about 150 K. At $P$ = 17 GPa, superconductivity is totally suppressed. Upon releasing pressure, the superconductivity is found to be reversibly restored (see Fig. \ref{RT} (c)). From the later Hall measurements as shown below, applying high pressure in fact acts to increase electron doping. No re-entrance of a second superconducting phase can be identified up to $P\sim$ 40 GPa (see Fig. \ref{RT} (b)). Besides, the AFM transition seen in NMR at the similar electron doping level does not seem to be reflected in the resistivity of our sample from the ambient pressure to the maximum pressure ($P \sim 40$ GPa). Compared with the previous results on Ca$_{1-x}$La$_x$FeAs$_2$ ($0.15 \leq x \leq 0.24$) \cite{Kawasaki} and Ca$_{1-x}$La$_x$Fe$_{1-y}$Co$_y$As$_2$ ($x= 0.27$) \cite{Jiang1} (see Fig. \ref{phasediagram}), the absence of AFM order in our studied sample may suggest the different doping mechanisms between the Ca-site doping and the Fe-site doping.

\begin{figure}
\includegraphics[width=8.0 cm]{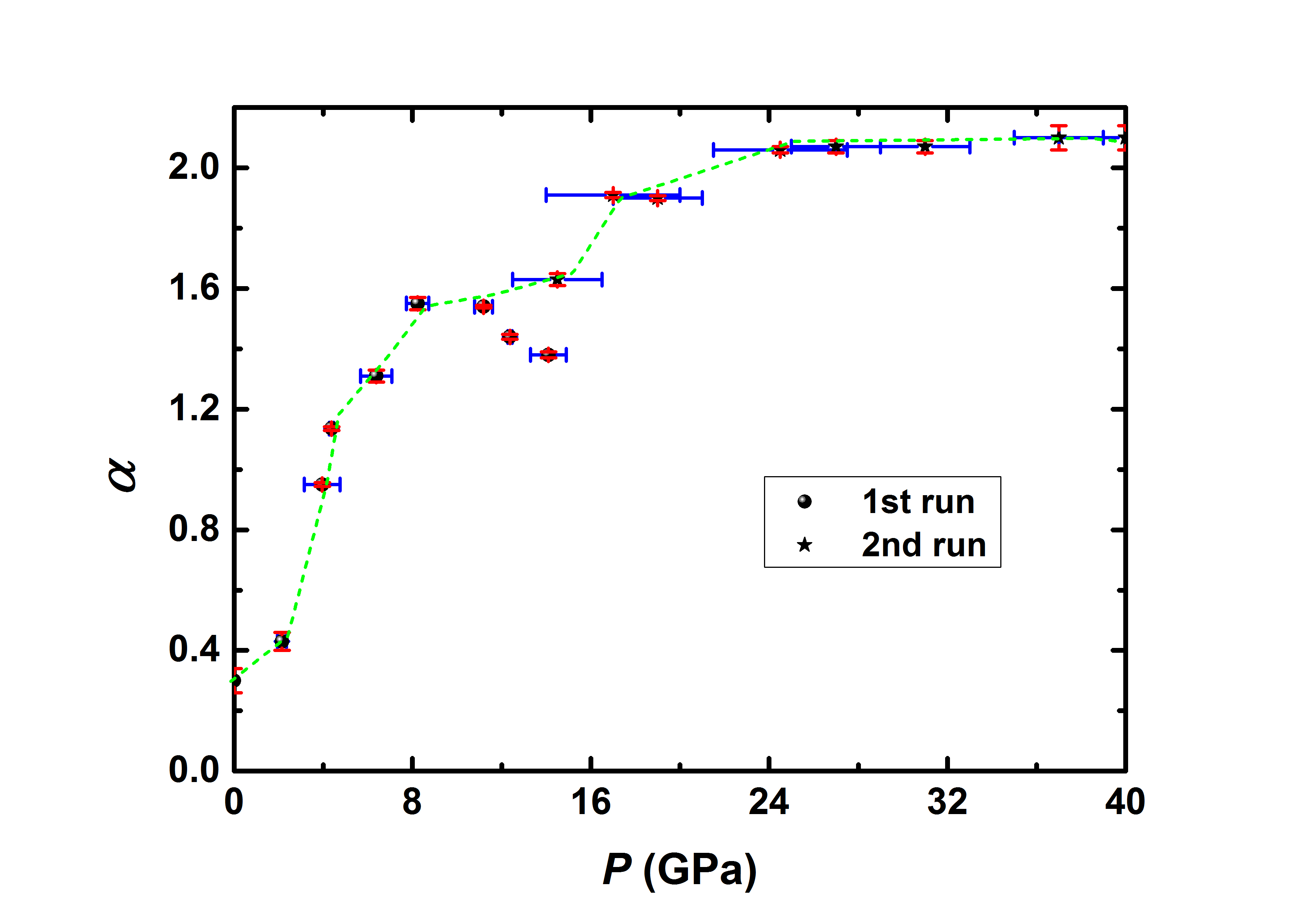}
\caption{\label{Exponent} Pressure dependence of the exponent $\alpha$ based on the power-law fitting in Fig. \ref{Powerlaw}. See the discussion on the large error bars of the second run in Ref. \cite{SI}.}
\end{figure}

In Fig. \ref{Powerlaw}, we further analyzed the normal state resistivity according to the power-law scaling, i.e., $\rho(T) = \rho_0 + AT^\alpha$, where $\alpha$ is the temperature exponent, $\rho_0$ is the residual resistivity, and the pre-factor $A$ is a parameter not only related to electron-electron interactions but also other interactions like electron-phonon coupling via the effective mass $m^\ast$. For a FL, $\alpha = 2$ and $A$ is determined by electron-electron scattering. As seen, at low temperatures, all $\rho(T)$ curves under different pressures nicely follow the power-law behaviors. For $P < 3.9$ GPa, the $\alpha$ values from the fitting are smaller than 1, indicative of substantial deviations from the FL behaviors. As $P$ increases, $\alpha$ value grows gradually. At $P = 3.9$ GPa, a linear $T$-dependence of $\rho$ is seen (inset in Fig.\ref{Powerlaw} (a)). It should be pointed out, strictly linear $\rho(T)$ has also been previously observed in other Fe-based systems like BaFe$_2$As$_{2-x}$P$_x$ \cite {Jiangshuai,Kasahara1} and BaFe$_{2-x}$Ru$_x$As$_2$ \cite{Eom}, which has attracted great attention due to the possible existence of a magnetic QCP. To date, whether there is a correlation between the linear $\rho(T)$ and a QCP is still an open question \cite{Analytis,Dai,HuD}. In this 112 system, the $T$-linear resistivity is likely a crossover from $\alpha <$ 1 to $\alpha >$ 1. As $P$ increases further, $\alpha$ value increases continuously until it saturates to $\alpha$$\sim$2 at $P = 17$ GPa, precisely where the superconductivity disappears. For $P \geq 17$ GPa, the low-temperature resistivity nicely obeys the $T^2$ rule, consistent with a FL ground state.

\begin{figure}
\includegraphics[width=8.0cm]{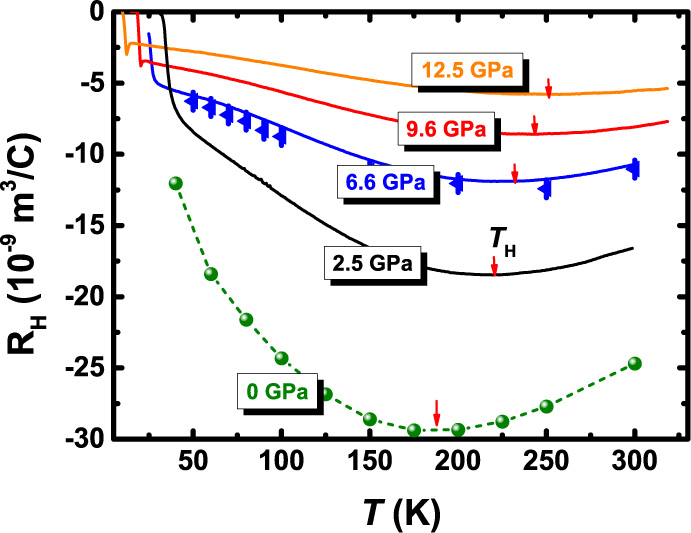}
\caption{\label{Hall} Temperature dependence of Hall coefficient $R_H$ under different pressures. The solid lines represent the measurements by $T$-sweeping method. The symbols stand for measurements by field-sweeping method. Both methods get consistent results. The temperature location of the extremum in $R_H(T)$ curve is labeled by $T_H$.}
\end{figure}

The $T$-dependence of exponent $\alpha$ is summarized in Fig. \ref{Exponent}. As seen, $\alpha$ keeps increasing from a value smaller than 1 at very low pressures to 2 for $P \geq$ 17 GPa. Here, we emphasize several distinct features by comparing the pressure effect of the present 112 system with the chemical doping effect in BaFe$_2$As$_{2-x}$P$_x$ or in BaFe$_{2-x}$Ru$_x$As$_2$. In the latter two cases \cite{Jiangshuai,Kasahara1,Eom}, AFM transition coexists with superconductivity in the under-doped region, and is gradually suppressed with increasing doping. At the optimally doped level, $\alpha$ is equal to 1. At the same time, AFM disappears and superconducting $T_c$ reaches its maximum. Below or above the optimal doping level, $\alpha$ is greater than 1. As a result, a V-shaped $\alpha(x)$ relation is observed. Here, $\alpha$ increases monotonously with $P$, and no adjacency to a \textit{static} AFM phase transition can be identified in the resistivity. Therefore no indication of the possible QCP can be identified in the present system. The residual resistivity $\rho_0$ and the pre-factor $A$ obtained through FL fitting above 17 GPa are shown in Ref. \cite{SI} (Fig. S3). Broadly speaking, in the framework of a FL theory, the pre-factor $A$ is proportional to the charge carrier effective mass $m^\ast$. The suppression of $A$ with $P$ suggests a reduced $m^\ast$, signifying gradual loss of electron correlations after superconductivity is suppressed.

\begin{figure}
\includegraphics[width=8.0cm]{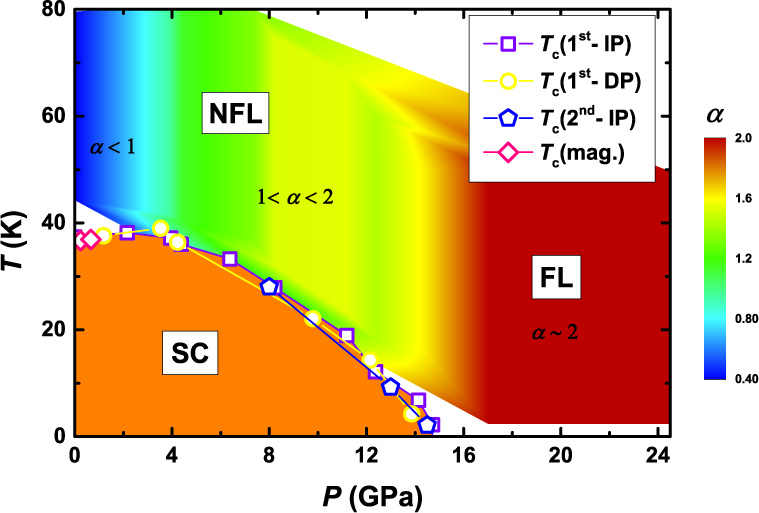}
\caption{\label{TcP} Pressure phase diagram. The evolution of the exponent $\alpha$ (indicated by color mapping) from the power-law fitting and the superconducting transition temperature $T_c$ versus pressure.}
\end{figure}

To proceed, we measured the Hall resistivity under different pressures. As seen from Fig. \ref{Hall}, the calculated Hall coefficients ($R_H$) from the field sweeping method and the temperature sweeping method show comparable values and consistent $T$-dependencies. At ambient pressure, $R_H$($T$) curve exhibits an obvious extremum at a characteristic temperature $T_H$ $\sim$ 180 K. Considering a single unit cell, the nominal doping carrier number is 0.22 per Fe atom based on the chemical stoichiometry of the sample. Then the nominal carrier density $n_e$ can be estimated to be $n_e = 0.22/V$, where $V$ is the volume of a single unit cell. Such a method leads to $|R_H| = 1/n_e \simeq 4.5*10^{-9}$ m$^3/$C. As can be seen from Fig. \ref{Hall}, this $|R_H|$ value is almost one order of magnitude lower than the experimental values. As pressure increases, while the characteristic $T_H$ of the $R_H$ extremum tends to move toward higher temperatures, the $R_H$ extremum seems to be wiped out. At the same time, the absolute values of $R_H$ decrease rapidly with increasing $P$, indicating a notable electron doping by pressure.

The above experimental results are summarized in the pressure-dependent phase diagram in Fig. \ref{TcP}. In this $T-P$ diagram, a superconducting dome is revealed with the pronounced normal-state NFL behaviors and monotonous $\alpha(P)$. Superconductivity disappears exactly when the NFL ground state is taken over by the FL. The simultaneous loss of $T_c$ and the NFL behaviors imply the same governing mechanism for both phenomena.

\begin{figure*}
\includegraphics[width=14cm]{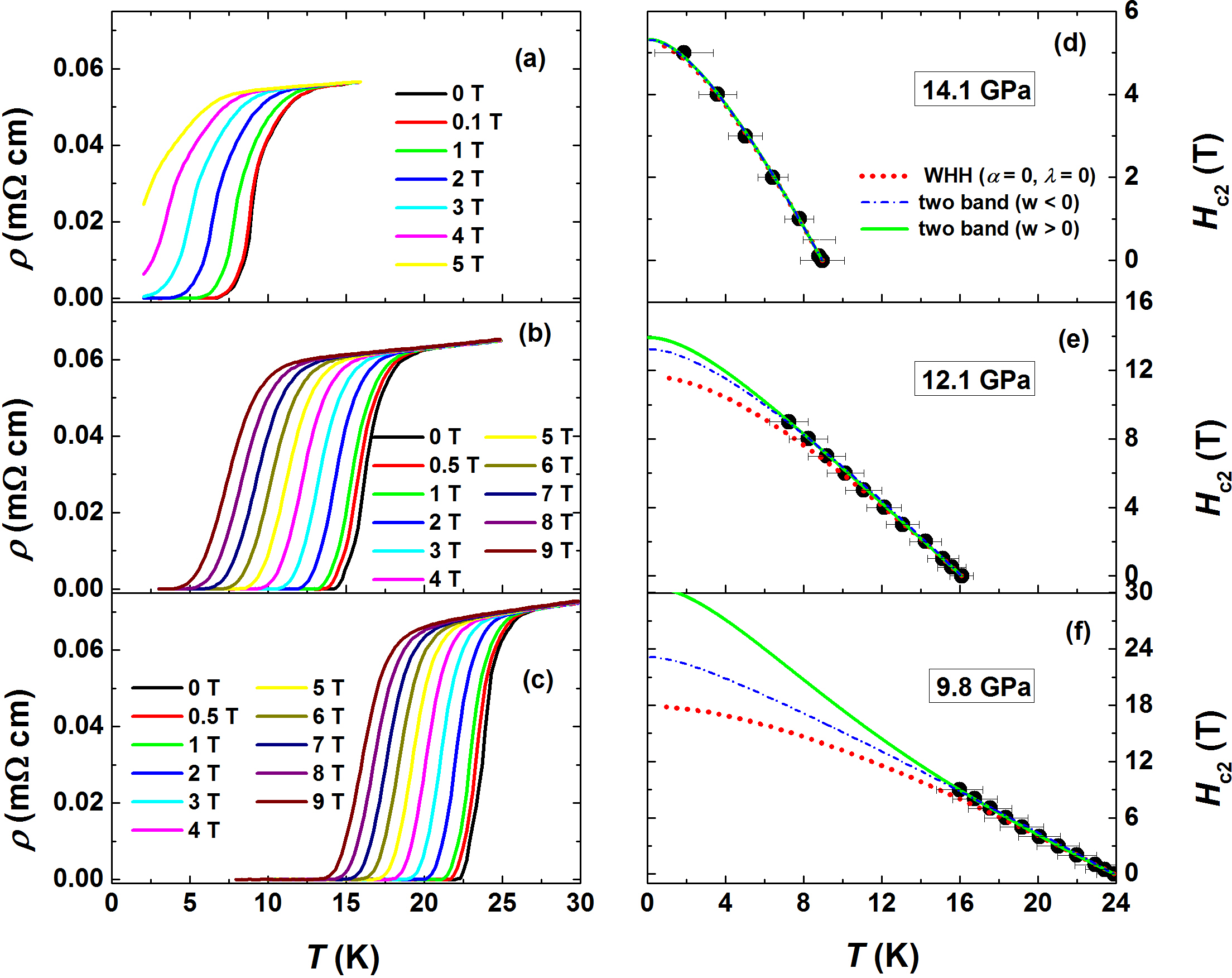}
\caption{\label{Hc2} (a-c) Temperature dependence of superconducting transitions under different magnetic fields applied along the $c$-axis for pressure $P$ = 14.1GPa, 12.1GPa, 9.8 GPa, respectively. (d-f) Temperature dependence of the upper critical field $H_{c2}$ which is extracted by the peak criterion, i.e., the peak position on $\frac{\partial \rho}{\partial T}(T)$ curves. The green solid lines and the blue dash-dot lines are two-band fittings for $H_{c2}(T)$ with $w > 0$ and $w < 0$, respectively. The red dot lines are WHH fits without considering spin-paramagnetic effect ($\alpha = 0$) and spin-orbit interaction ($\lambda = 0$). The error bars are determined by half of the difference of $H_{c2}(T)$ between the peak criterion and the 90\% criterion, which must be larger than the maximum fluctuation of actual $H_{c2}(T)$. The 90\% criterion defines $H_{c2}$ by 90\% of the normal-state resistivity just above $T_c$.}
\end{figure*}

\begin{table*}
\caption{\label{Hc2fitting}Parameters obtained from the WHH and two-band fittings. $H_{c2}^{\text{WHH}}$ is from WHH fitting, and $H_{c2}^{\text{two}}$ is from two-band fitting. $w = \lambda_{11}\lambda_{22}-\lambda_{12}\lambda_{21}$.}
\begin{ruledtabular}
\begin{tabular}{cccccccccc}
$P$&$T_c$(0 T)
&$-dH_{c2}/dT|_{T_c}$
&$H_{c2}^{\text{WHH}}$(0 K)&& $w > 0$ & & & $w < 0$ & \\
(GPa)&(K)&(T/K)&(T)&$H_{c2}^{\text{two}}$(0 K) (T)& $D_1$ &$\eta$ ($=D_2$/$D_1$)&$H_{c2}^{\text{two}}$(0 K) (T)& $D_1$ &$\eta$ ($=D_2$/$D_1$)\\
\hline
14.1 & 8.95 & 0.85& 5.24 & 5.31 & 1.28 &1 & 5.31 & 1.35 &0.9 \\
12.1 & 16.08 & 1.05 & 11.81 &13.92 & 1.5 & 0.4 & 13.25 & 1.45 &0.37\\
9.8  & 23.89 & 1.08 & 17.79 & 30.61 & 1.8 & 0.18 & 23.18 & 1.62 &0.18\\
0$^{\text{Ref.}}$\cite{Xingxz} & 38.8 & 2.26 & 61 & 88.34 & 2.4 & 0.13 &/ & /&/ \\
\end{tabular}
\end{ruledtabular}
\end{table*}

We further investigate the evolution of the upper critical field $H_{c2}$ ($H$ $\parallel$ $c$-axis) under pressure. In Fig. \ref{Hc2}, the superconducting transitions under different magnetic fields and the temperature dependence of $H_{c2}$ for different pressures are shown. In Fe-based superconductors, it is generally perceived that the single band Werthamer-Helfand-Hohenberg (WHH) theory involved with spin-paramagnetic effect via the Maki parameter $\alpha_M$ and the spin-orbit interaction constant $\lambda$ can overall fit the experimental $H_{c2}$ for $H\parallel ab$, while for $H_{c2}$ along the $c$-axis the two-band model has to be considered \cite{AGurevich,Zocco,Wangzhaosheng}. In the absence of both spin-paramagnetic effect ($\alpha_M = 0$) and spin-orbit interaction ($\lambda = 0$), WHH formula can be simplified as \cite{Werthamer}

\begin{equation}
\ln\frac{1}{t}=\sum^{\infty}_{\nu=-\infty}\left\{\frac{1}{|2\nu+1|}-[|2\nu+1|+\frac{\overline{h}}{t}]\right\},
\end{equation} where $t=T/T_c$ and $\overline{h}=(4/\pi^2)[\frac{H_{c2}(T)}{(\frac{dH_{c2}}{dt})}|_{T_c}]$. At $P = 14.1$ GPa, as seen, a WHH fit with $\alpha = 0, \lambda = 0$ can reproduce the experimental data very well. However, at lower pressures, WHH fitting curves slightly fall below the experimental $H_{c2}$ data. Alternatively, a two-band description of $H_{c2}(T)$ as in MgB$_2$ should be applied. The two-band model is expressed as \cite{Gurevich}

\begin{widetext}
\begin{equation}
a_0[\ln t+U(h)][\ln t+U(\eta h)]+a_2[\ln t+U(\eta h)]+a_1[\ln t+U(h)]=0.
\end{equation}
\end{widetext}
Here, $a_0$, $a_1$, $a_2$ are determined by $\lambda$ matrix
$\left(
  \begin{array}{cc}
    \lambda_{11} & \lambda_{12}\\
    \lambda_{21} & \lambda_{22}\\
  \end{array}
\right),$ where $\lambda_{11}$, $\lambda_{22}$ and $\lambda_{12}$, $\lambda_{21}$ are intraband and interband coupling constants. $U(x)=\psi(1/2+x)-\psi(1/2)$, where $\psi(x)$ is the digamma function. $h=H_{c2}D_1/2\phi_0T$ and the band diffusivity ratio $\eta=D_2/D_1$, where $\phi_0$ is the flux quantum and $D_1$, $D_2$ are diffusivity of different bands. We first assume an intraband-dominant coupling with $\lambda=\left(
  \begin{array}{cc}
    0.5 & 0.25\\
    0.25 & 0.5\\
  \end{array}
\right)$, i.e. $w = \lambda_{11}\lambda_{22}-\lambda_{12}\lambda_{21} > 0$, which is often used for Fe-based superconductors. Then, we only tune $D_1$ and $\eta$ to fit the experimental data. As seen from Figs. \ref{Hc2} (d)-(f), the modeling of $H_{c2}(T)$ based on the two-band model is clearly better in comparison with the WHH fits for pressure $P = 9.8 $GPa and 12.1 GPa. We further adopt an interband-dominant coupling with $\lambda=\left(
  \begin{array}{cc}
    0.25 & 0.5\\
    0.5 & 0.25\\
  \end{array}
\right)$, i.e. $w = \lambda_{11}\lambda_{22}-\lambda_{12}\lambda_{21} < 0$, to fit the experimental $H_{c2}(T)$. As can be seen, the fittings are also reasonable, and the values of $D_1$ and $\eta$ versus $P$ develop the same trend as the intraband-dominant case. For $P = 14.1$ GPa, the obtained $\eta$ values for both cases are around 1. $\eta$=1 indicates equal band diffusivities for the two bands, which in fact reduces to the single-band WHH model in the dirty limit. The WHH and two-band fitting parameters are listed in Table I for completeness. Note that our analysis is rather independent of criteria used in determining $H_{c2}$. Here we define $H_{c2}$ as the peak in $\frac{\partial \rho}{\partial T}(T)$ curves under individual fields. We also use the criterion of 90\% of the normal state $\rho_n$ (see Ref. \cite{SI}) and get the same results.

\section{Discussions and Conclusion}

A strong temperature dependence of $R_H$, in particular the extremum around 180 K at ambient pressure is striking, which was seen in some other Fe-based superconductors, such as the 122 pnictides as well as LiFeAs and LiFeP \cite{Kasahara,Ohgushi}. Many theoretical works \textit{all} emphasize the anisotropic interband scattering between electronlike and holelike Fermi surfaces as the possible origin for this Hall coefficient extremum \cite{Fanfarillo,Breitkreiz,Breitkreiz1}. The most likely source for this anisotropic scattering is thought to come from the SFs, which induce a mixing of electron and hole currents such that the renormalized current in each band can even possesses an opposite direction (negative transport time) with respect to the bare band velocity. In this picture, $|R_H|$ extremum tends to be high for higher scattering anisotropy between electron ($e$) and hole ($h$) bands. The suppression of the Hall extremum under pressure seems to imply the decrease of anisotropic scattering and hence the weakening of SFs with pressure. Indeed, the suppression of $|R_H|$ extremum with increasing $P$ is also compatible with the $T_c$ decrease, which favors the widely perceived pairing picture for iron pnictides, namely, SFs mediated pairing. This in turn supports that the NFL in the normal state is also induced by SFs.

As we found, the downward curvature ($\alpha < 1$) in $\rho(T)$ curves at low pressures was also reported in the high-$T$ phase of over-doped Ba$_{1-x}$K$_x$Fe$_2$As$_2$ superconductors \cite{Shen,Ohgushi,Liu}, where a crossover from high-$T$ incoherent state ($\alpha < 1$) to low-$T$ coherent state ($\alpha > 1$) was claimed to be seen. In our system, the suppression of the incoherent state with pressure up to optimal pressure (3.9 GPa) seems to explain the initial increase of $T_c$ under pressure.

Last, the pressure dependence of $H_{c2}$ is also in favor of the above SFs picture. At low pressures, SFs are strong, which favors the strong anisotropic superconducting gap or two gaps in the electronic spectrum. Our previous high-field experiments have given strong evidence for the two-band effects for the same sample at ambient pressure \cite{Xingxz}. As pressure increases, SFs become effectively reduced and the band diffusivity ratio $\eta=D_2/D_1$ increases toward 1. $\eta =1$ suggests more similar intraband scattering in the electron ($e$) and the hole ($h$) Fermi sheets, which ultimately leads to the thermodynamically isotropic gap (one gap). Overall, the above pressure dependence of $T_c$, $R_H$, $H_{c2}$, as well as the NFL to FL crossover all seems to support the picture that the SFs are suppressed by pressure and the superconductivity in this system is indeed mediated by SFs.

In summary, we have investigated the superconductivity and the normal state transport behaviors in a 112-type superconductor Ca$_{1-x}$La$_x$Fe$_{1-y}$Co$_y$As$_2$ ($x= 0.2$, $y= 0.02$) in a broad pressure range up to 40 GPa. In the $T-P$ phase diagram, it is found, the normal-state of the whole superconducting region is accompanied with the non-Fermi-liquid behaviors which turn into Fermi-liquid when superconductivity is totally suppressed. Exceptionally, no evidence of static AFM order is observed by transport measurement accompanying the non-Fermi-liquid region. Detailed study on the pressure evolution of the notable $R_H$ extremum and the two-band nature of $H_{c2}$ suggests non-negligible influence from the anisotropic scattering between electron- and hole- bands which is presumably caused by the spin fluctuations.

\begin{acknowledgments}
We appreciated the fruitful discussions with Maxim Breitkreiz and Carsten Timm, and the careful proofreading by Ali Bangura. This work was supported by the National Natural Science Foundation of China (Grant No. U1432135, 11611140101, and 11674054). X. X. acknowledges the financial support from the National Key Basic Research Program of China (Grant No. 2014CB648400) and from National Natural Science Foundation of China (Grant No. 11474080, No. U1732162). B. Q. was supported partly by the Natural Science Foundation of Jiangsu Province (Grant No. BK20150831), Natural Science Foundation of Jiangsu Educational Department (Grant No. 15KJA430001), and six-talent peak of Jiangsu Province (Grants No. 2012-XCL-036). W. Zhou and F. Ke contributed equally to this work.
\end{acknowledgments}

\end{document}